\documentclass{aa}  

\usepackage{txfonts}
\usepackage[colorlinks=true,linkcolor=blue, citecolor=blue, filecolor=blue, urlcolor=blue]{hyperref}
\begin{document}

   \title{Exploring non-thermal emission from the star-forming region NGC~3603 through a realistic modelling of its environment}
    \titlerunning{NGC 3603 non-thermal emission}

   \author{M. Rocamora\inst{1}, A. Reimer\inst{1}, G. Martí-Devesa\inst{2,3} \and R. Kissmann\inst{1}}

   \institute{Universität Innsbruck, Institut für Astro- und Teilchenphysik, Technikerstr. 25/8, 6020 Innsbruck, Austria
   \\\email{manuel.rocamora@uibk.ac.at}
   \and
         Dipartimento di Fisica, Universit\'a di Trieste, I-34127 Trieste, Italy
   		\and
 		Istituto Nazionale di Fisica Nucleare, Sezione di Trieste, 34127 Trieste, Italy\\
   }

   \date{Received September 15, 1996; accepted March 16, 1997}

 
  \abstract
   {Star-forming regions are gaining considerable interest in the high-energy astrophysics community as possible Galactic particle accelerators. In general, the role of electrons has not been fully considered in this kind of cosmic-ray source. However, the intense radiation fields inside these regions might make electrons significant gamma-ray contributors.}
   {We study the young and compact star-forming region NGC 3603, a well known gamma-ray emitter. Our intention is to test whether its gamma-ray emission can be produced by cosmic-ray electrons.}
   {We build a novel model by creating realistic 3D distributions of the gas and the radiation field in the region. We introduce these models into PICARD to perform cosmic-ray transport simulations and produce gamma-ray emission maps. The results are compared with a dedicated \textit{Fermi} Large Area Telescope data analysis at high energies. We also explore the radio and neutrino emissions of the system.}
   {We improve the existing upper limits of the NGC 3603 gamma-ray source extension. Although the gamma-ray spectrum is well reproduced with the injection of CR protons, it requires nearly 30\% acceleration efficiency. In addition, the resulting extension of the simulated hadronic source is in mild tension with the extension data upper limit. The radio data disfavours the lepton-only scenario. Finally, combining both populations, the results are consistent with all observables, although the exact contributions are ambiguous.}
   {}


   \maketitle
%
\nolinenumbers
\section{Introduction}

    The origin of cosmic rays has been an intriguing question since their discovery in the early 20th century. The most accepted scenario for Galactic cosmic rays (CR) places their origin in supernova remnants (SNRs). However, this scenario has difficulties to explain some aspects of the available data \citep{Gabici_2019}. For example, the maximum energy usually attributed to Galactic cosmic rays (i.e. a few PeV) is only achievable in SNRs with specific characteristics and having highly enhanced magnetic fields \citep{Lagage_1983, Bell_2004, Cristofari_2020}. Another problem in this standard scenario is the chemical composition of cosmic rays. The SNR origin cannot account for the overabundance of some isotopes measured in the cosmic-ray spectrum at Earth, such as $^{22}$Ne, which is only significantly produced in Wolf-Rayet (WR) stars \citep{Higdon_2003}.
    These problems have motivated searches for alternative cosmic-ray sources within the Galaxy that can help to better fit the data.

    Young and compact star-forming regions were long ago proposed as possible cosmic-ray sources \citep{Cesarsky_1983}, but they were not taken into full consideration until the above-mentioned problems arose. These regions typically contain many massive stars, which lose material through powerful winds and form a cavity in their surrounding medium \citep{Weaver_1977, Balick_1980,Gallegos_2020}. In this cavity, also known as bubble, the collective winds of the stars develop a termination shock where particles can accelerate through diffusive shock acceleration \citep[DSA;][]{Bykov_2014, Morlino_2021}. 
    
    Star-forming regions also have some difficulties to reach PeV energies \citep{Morlino_2021, Vieu_2022b}, mainly due to the size of the acceleration region. Nevertheless, they can easily solve the composition problem \citep{Binns_2008,Gupta_2020}, since WR stars usually dominate the winds in these high-energy sources \citep[see e.g.][]{Menchiari_2024}. The number of detected systems in the gamma-ray band is rising thanks to technical improvements and the accumulation of data \citep[e.g.][]{Ackermann_2011,Abramowski_2012,Abramowski_2015,Yang_2018,Abeysekara_2021}, increasing the interest in understanding the acceleration mechanisms that might be playing a role in there. For example, it is not clear whether their cosmic-ray content is dominantly hadronic or leptonic: There have been many studies regarding the acceleration of particles in these systems where the leptonic component is, in general, disregarded \citep{Maurin_2016,Aharonian_2019,Bykov_2020,Vieu_2022a,Bykov_2022}. However, the recent Westerlund 1 measurements by H.E.S.S. \citep{Aharonian_2022} seem to be better explained using a leptonic scenario \citep{Harer_2023} for the gamma-ray origin.

    In this work, we explore the non-thermal properties of NGC 3603, a young and dense star-forming region. This system has been deeply studied over the whole electromagnetic spectrum and its physical characteristics are well determined. At a distance of $\sim$ 7 kpc \citep{Moffat_1983}, the region contains many young massive OB stars \citep{Moffat_1994,Sung_2004}. Furthermore, a gamma-ray source in the direction of NGC 3603 was already detected \citep{Yang_2017,Saha_2020}, but the particle population underlying this emission is still unclear. It is also intriguing that, even though it is one of the most dense and powerful star-forming regions in the Galaxy, hence ideal for particle acceleration, its detection has not been reported by TeV instruments.
    
    Our aim is to probe whether it is possible that gamma-ray emission from NGC 3603 is produced by cosmic-ray electrons accelerated by the star cluster. For this purpose, we build a model with realistic gas density and radiation field distributions, while previous attempts employed mean values for the whole region. This realistic modelling, based on multiwavelength observations, allows us to study more realistically the non-thermal morphology of the system. 

    We make use of the PICARD code \citep{Kissmann_2014}, a cosmic-ray propagation solver usually applied to the Milky Way. We introduce our 3D distributions of the gas density and radiation fields in PICARD; inject and propagate the CR particles; calculate their gamma-ray products and, finally, compare with \textit{Fermi} Large Area Telescope (\textit{Fermi}-LAT) data.
    We re-analysed this region using 5 more years of data with respect to the last analysis of Fermi data \citep{Saha_2020} and optimize the selection cuts for the morphological study.
    In addition, we also explore the radio and neutrino emissions of our scenarios, as more observables add further constraints to our models. 
    
    The paper is organised as follows: our model is detailed in Section~\ref{sec:model}; in Section~\ref{sec:analysis}, we describe our analysis of \textit{Fermi}-LAT gamma-ray data; we show our results in Section~\ref{sec:results}; and we finish with a brief discussion and conclusions in Sections~\ref{sec:discussion} and \ref{sec:conclusion}, respectively.


\section{Model}
\label{sec:model}
    To compute the injection and propagation of cosmic rays within the star-forming region, we make use of the PICARD code to solve the diffusion-loss equation \citep{Ginzburg_64, Berezinskii_90}
    \begin{align}
        \nonumber 
        -\overrightarrow{\nabla}\cdot\left(D\overrightarrow{\nabla}\Psi_i - \overrightarrow{v}_w\Psi_i\right) + \frac{\partial}{\partial p}\left[p^2 D_{pp}\frac{\partial}{\partial p} \left(\frac{\Psi_i}{p^2}\right)\right] - \\ 
        \nonumber - \frac{\partial}{\partial p} \left[\dot{p}\Psi_i - \frac{p}{3}\left(\overrightarrow{\nabla}\cdot \overrightarrow{v}_w\right)\Psi_i\right] = Q \, + \\ +\sum_{i<j}\left(c\beta n\sigma_{j\rightarrow i} + \frac{1}{\gamma \tau_{j\rightarrow i}}\right)\Psi_j - \left(c\beta n\sigma_i + \frac{1}{\gamma\tau_i}\right)\Psi_i
        \label{eq:prop}
    \end{align}
    where $D$ is the spatial diffusion coefficient; $\Psi$ represents the cosmic-ray flux; $v_W$ is the wind velocity; $p$ the momentum of the particles, $D_{pp}$ is the momentum diffusion coefficient; $Q$ the source term; $c$ the speed of light; $\beta c$ and $\gamma$ are the velocity and Lorentz factor of the particles, respectively; $n$ is the gas density; $\sigma_{j\rightarrow i}$ and $\sigma_i$ are the production and fragmentation-loss cross-section, respectively; and, finally, $\tau$ is the decay mean lifetime of the species. This equation is solved in a 3D Cartesian grid of 257$\times$257$\times$257 cells, with x, y, z $\in$ [-64, 64] pc. The boundary condition is that the CR flux vanishes at the edge of the simulation box. The box size is chosen making sure that this boundary condition does not affect the results. The central cell in the box represents the sky position RA = 11h 15' 05.3'', Dec = -61$^\circ$ 15' 43'' at a distance d = 7~kpc. The energy range considered is E $\in$ [100 MeV, 1 PeV], with 127 logarithmically-spaced bins.

    To solve the propagation equation we need to implement the astrophysical environment of the star-forming region into PICARD, as gas and photon field distributions impact CR propagation. 

\subsection{Radiation field}
    
    In order to model a realistic radiation field, we make use of a catalogue of more than 200 O stars within the NGC 3603 region and its surroundings \citep{Drew_2019}, since our simulation box is bigger than the SFR itself. It contains the effective temperature and spatial coordinates for every star. As these will be the main source of photons within the cluster, we neglect the contribution of less massive stars.

    \begin{figure}
        \centering
        \includegraphics[width=\hsize]{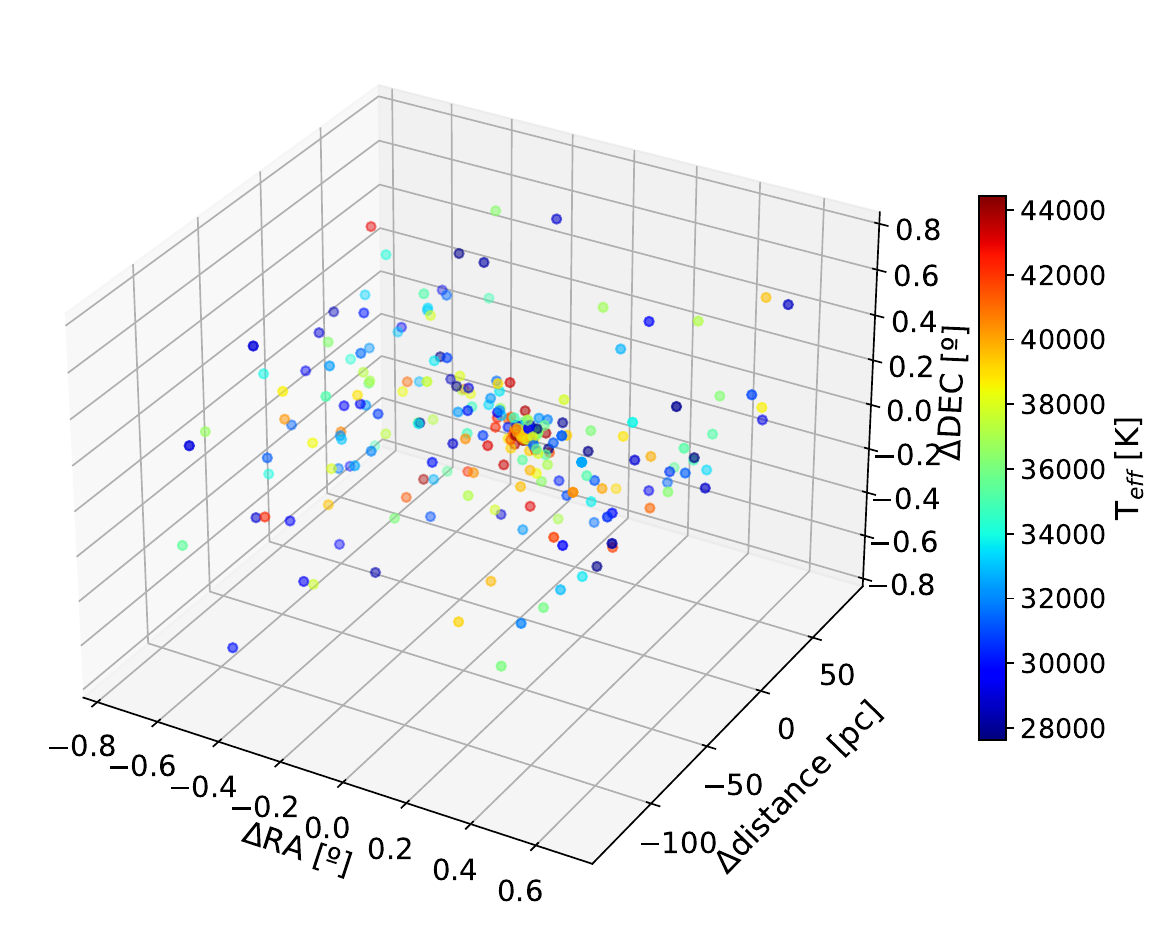}
        \includegraphics[width=0.95\hsize]{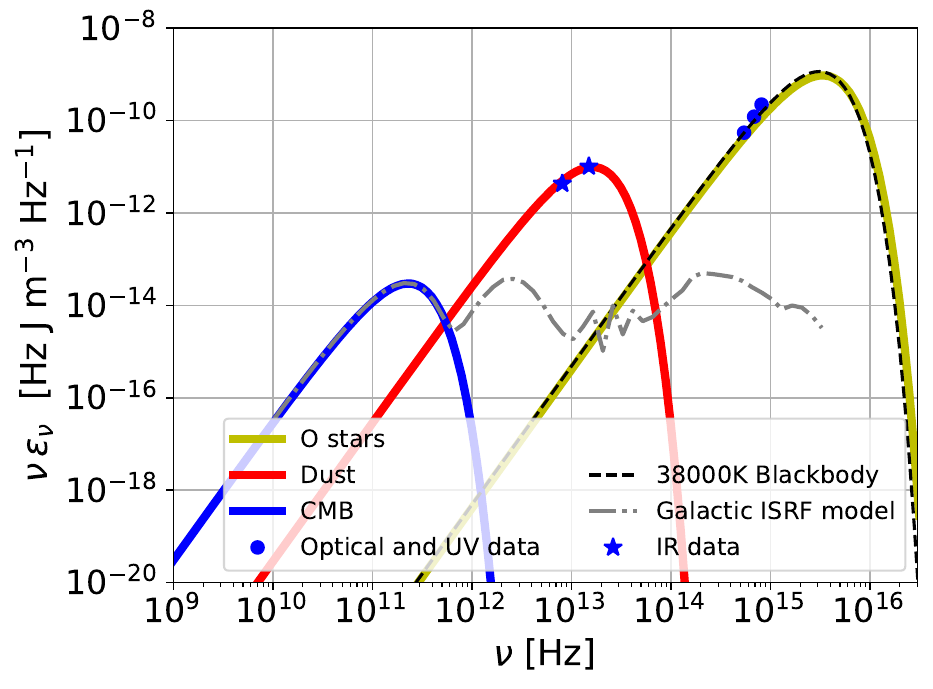}
        \caption{3D distribution of O stars in our model (top) and radiation field density at the center of the simulation box (bottom). Colours in the top panel represent the effective temperature of the stars. The blue data points in the lower panel are from \cite{Sher_1965}, stars show the data from \cite{DeBuizer_2024}, while the Galactic ISRF model is from \cite{Porter_2005}.}
        \label{fig:stars}
    \end{figure}
    
    To calculate the radiation density at each point, we need to place the stars inside our 3D model. However, since NGC 3603 is far away, the exact distances of the stars are poorly constrained \citep{Drew_2019}. We take the position of the stars in the sky and calculate their projected distance to the center of the star cluster. With the standard deviation of this quantity, which is 21 pc, we generate a Gaussian distribution centered at 7 kpc. We use this distribution to set randomly the distance of every single star. The resulting distribution of stars is shown in the upper panel of Figure~\ref{fig:stars}.

    We then assume that the stars emit as a black body and the contribution of every single star decreases as the distance squared. In addition, we also include the Cosmic Microwave Background (CMB) radiation and a dust component. The resulting radiation density spectrum is shown in the lower panel of Figure~\ref{fig:stars}. The contribution of all the stars resembles a single black body with a temperature of 38000 K. The data points show the radiation inside the star-forming region estimated from infrared, optical and ultraviolet data \citep{DeBuizer_2024,Sher_1965} and applying the corresponding extinction correction factor for each band \citep{Pang_2016}. There is a difference of around 3-4 orders of magnitude between our model and typical Galactic Interstellar Radiation Field (ISRF) models \citep[e.g.]{Porter_2005} in that region. Such difference might be due to the axial symmetry of the ISRF models, which averages over the azimuth angle of the Galaxy.

    \begin{figure}
        \centering
        \includegraphics[width=\hsize]{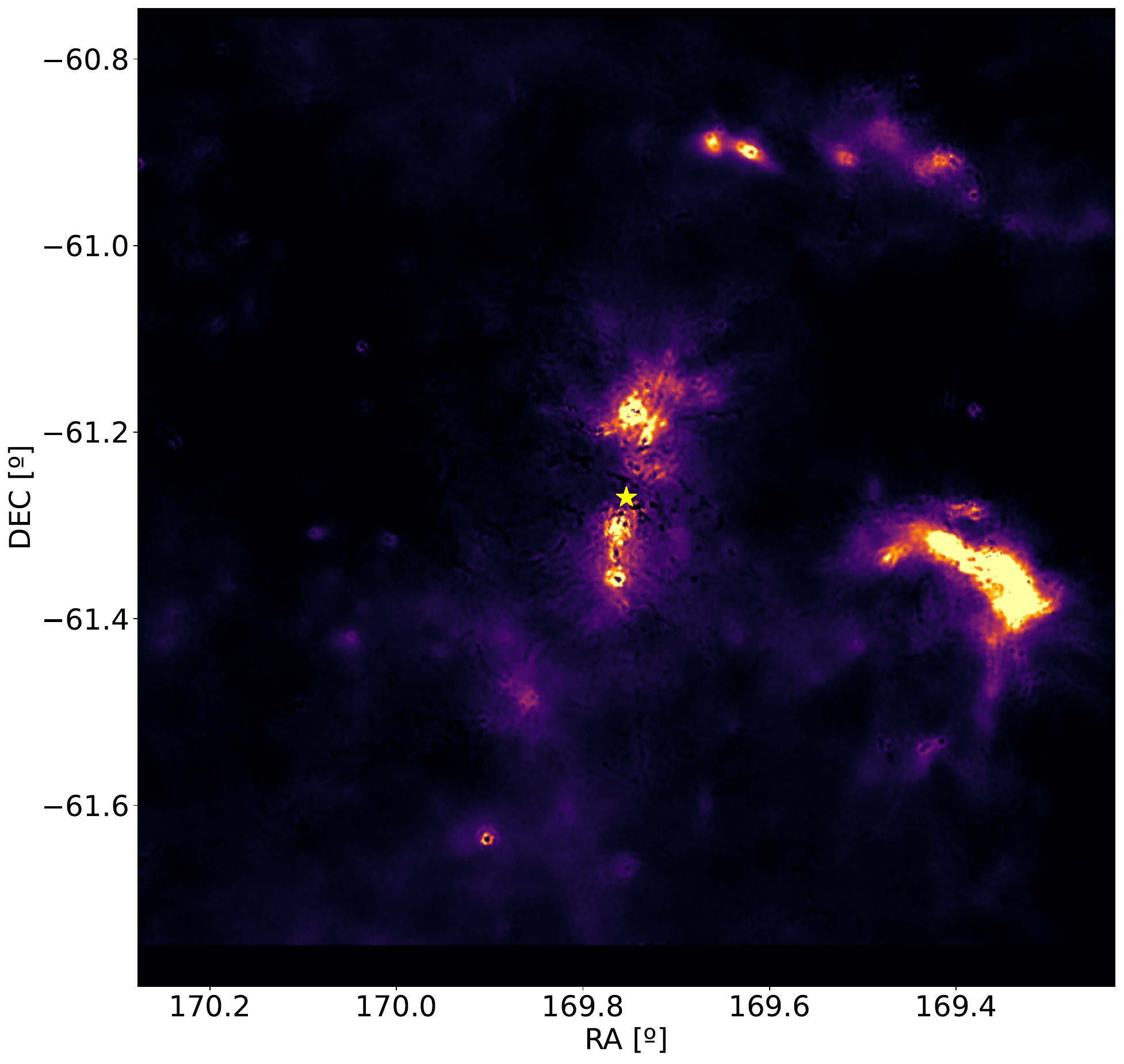}
        \includegraphics[width=\hsize]{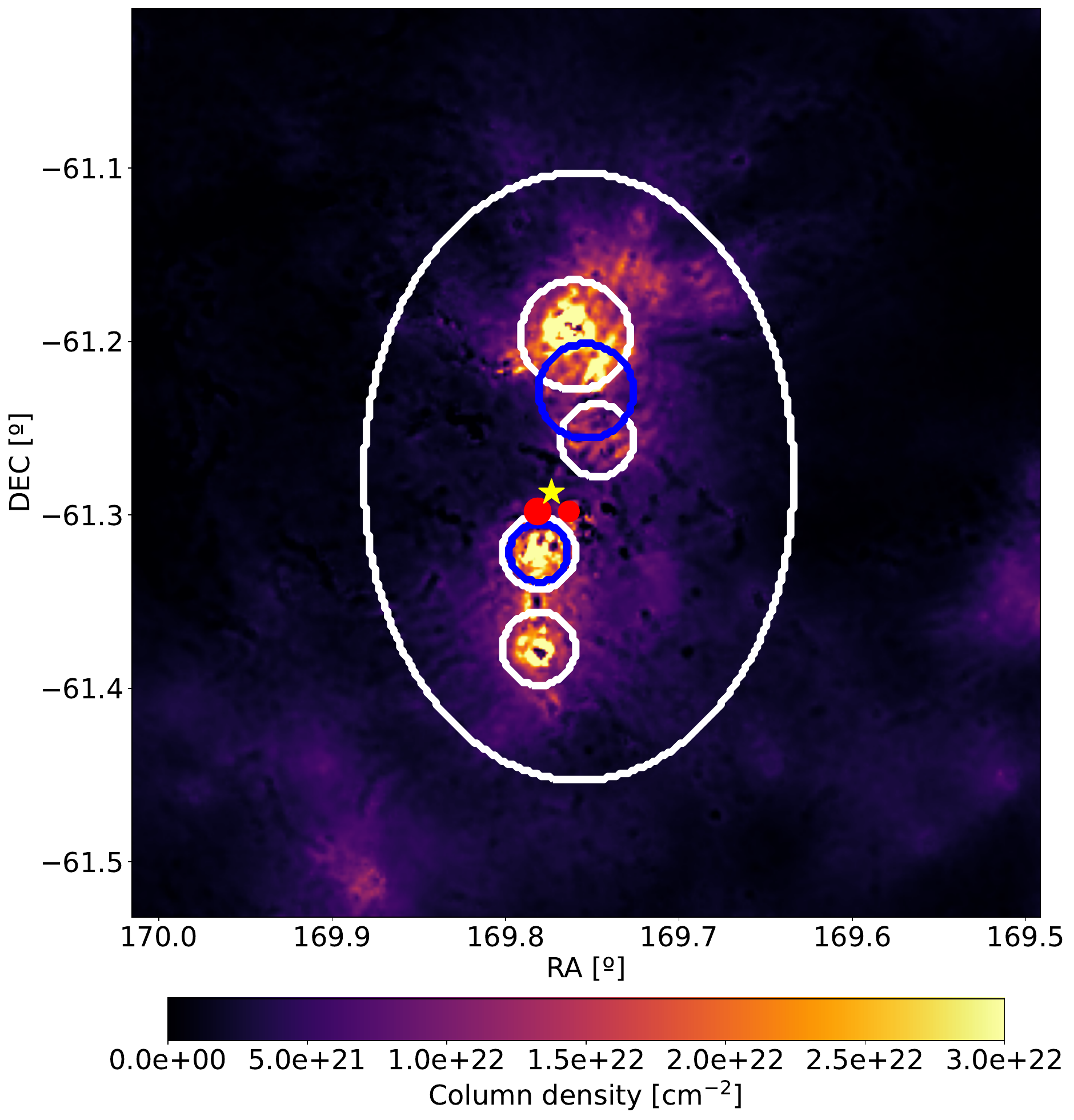}
        \caption{Molecular gas column density in the NGC 3603 region (top) and zoomed central region (bottom). The yellow star marks the position of the stellar cluster. The size of the upper map represents the size of the simulation box. White circles and ellipses show the molecular clouds introduced in our simulation box, blue circles show the position of the atomic clouds and, finally, filled red circles show the ionized gas.}
        \label{fig:H2}
    \end{figure}

\subsection{Gas density}

    \begin{table*}
        \centering
        \begin{tabular}{|c|ccccccccc|}
        \hline
        Cloud & H$_2$ 1 & H$_2$ 2 & H$_2$ 3 & H$_2$ 4 & H$_2$ 5 & HI 1 & HI 2 & HII 1 & HII 2\\ \hline
        Radius [pc] & 3.7 & 2.4 & 2.5 & 2.4 & 7.3, 10.2 & 3.2 & 2 & 0.7 & 1\\
        $\rho$ [cm$^{-3}$] & 1750 & 1000 & 2000 & 1780 & 60 & 440 & 1040 & 80 & 50\\ \hline
        \end{tabular}
        \caption{Properties of the clouds included in the simulation. The values presented as radius for the cloud H$_{2}$ 5 are the minor and major axis of the ellipsoid.}
        \label{tab:clouds}
    \end{table*}
    
    The main gas density component in the region is the molecular gas. We follow \cite{Schneider_2015} and perform a graybody fit to Herschel data in order to calculate the H$_2$ column density. Our results of the fit are shown in Figure~\ref{fig:H2}. Although a bit noisier, they are similar to the original work. This noise will be averaged out when implemented into PICARD due to the spatial resolution of the simulation.
    
    The column density is the integrated gas density along the line of sight in the region. To determine its 3D distribution, we assume that the clouds have constant density and their centroids are located at the same distance as the star-forming region. The clouds that are introduced in the simulation box are marked with white circles and ellipses in Figure~\ref{fig:H2}. There are four small dense clouds and a bigger elliptical one in order to account for the leftovers. Their densities and radii are shown in Table~\ref{tab:clouds}. 

    It is not clear whether the molecular cloud located at RA 169.4$^\circ$ Dec -61.35$^\circ$ belongs to NGC 3603. It could be related to the star-forming region NGC 3756, which is located to the west of our studied region, but at a distance of 2.8 kpc \citep{dePree_1999}. We use the MOPRA survey of CO velocity maps \citep{MOPRA} to verify this. While the molecular clouds belonging to NGC 3603 are found between 8 and 22 km/s \citep{Rollig_2011}, the outer cloud is in the range -30 to -16 km/s. Thus, this cloud appears to be not linked to NGC 3603, and is therefore not included in our simulation.

    For atomic hydrogen, no detailed data is available. Hence we use the resulting constant density and cloud size from \cite{Retallack_1980}, based on observations at 1415 MHz (21 cm). These are two clouds, which are shown as blue circles in Figure~\ref{fig:H2}. Their physical characteristics are also described in Table~\ref{tab:clouds}. Finally, there are two main clumps of ionized hydrogen, as seen in \cite{Nurnberger_2003}, shown as red circles in our figure. In this case, we use the electron population from \cite{McLeod_2016}, based on optical Multi-Unit Spectroscopic Explorer (MUSE) data. In addition, we take into account Helium in all our target gas components, with a Helium-to-Hydrogen ratio of 0.11.

\subsection{Particle injection and propagation}

    To understand the origin of the gamma-ray emission, we consider three different scenarios for the particle injection: a purely leptonic case, where we inject only electrons; a purely hadronic scenario, where we inject only protons; and, finally, a hybrid setup, where both electrons and protons are injected. Additionally, secondary CR electrons and positrons will also be produced in the scenarios containing CR protons through pp-interactions.
    
    We assume the acceleration mechanism in the system to be DSA at the termination shock. We do not implement the build-up of the spectrum of accelerated particles by the DSA mechanism, but  simply inject the readily accelerated particles at the shock position. To locate the termination shock, we use a spherical shell with half of the bubble radius. The bubble evolution equations \citep{Weaver_1977}, in the case of NGC 3603, predict a radius of
    \begin{equation}
        R_{b} = L_w^{1/5} n_{ext}^{-1/5} t^{3/5} \approx 80~\mathrm{pc},
    \end{equation}
    with $n_{ext}$ = 10~cm$^{-3}$ and $L_W\approx3.2\cdot10^{38}$~erg/s \citep{Drissen_1995}. This kinetic wind luminosity value is calculated summing the mass outflows of the most massive stars within NGC 3603 and is dominated by three WR systems, whose contribution represents around 60\% of the total stellar wind power \citep{Crowther98}. Observations, however, have revealed a value of $R_{b}\sim 1$~pc \citep{Balick_1980}, much smaller than the theoretical prediction. This might be due to the high-density clouds surrounding the star-forming region or because the stars were not born at the same time. In this study we proceed with the observational value, which leads to a radius for the termination shock $R_{ts}=R_b/2 \approx$ 0.5 pc. 
    
    For the energy dependence of the particle injection we use a single power-law with an exponential cutoff:
    \begin{equation}
        Q(E,r) = Q_0 \,E^{-\alpha} \exp(-E/E_c)~\delta(r-R_{ts}),
    \end{equation}
    where $Q_0$ is given by demanding that the injected energy is a fraction of the wind kinetic luminosity
    \begin{equation}
        \eta\,L_W = \int_{V} \int_{E_0}^{E_1} \frac{4\pi}{\beta c} EQ(E)~ dE dV,
    \end{equation}
    where $\eta$ is the efficiency of the acceleration. 

    After being injected, the particles propagate and suffer further energy gains, through second order Fermi acceleration, and losses. The following energy-loss processes are included in Eq.~\ref{eq:prop}: pion production, Coulomb scattering, ionization, bremsstrahlung, synchrotron radiation, and inverse Compton (IC) scattering. The maximum energy achievable by electrons is determined by balancing the sum of all loss rates with the acceleration rate $\tau_{acc} \approx V_w^2/8D$. 
    This results in an electron cutoff energy of $E_{c,e}=10$ TeV. On the other hand, the energy losses are negligible for protons and the corresponding spectral cutoff energy is given by the size of the acceleration region through $E_{c,p}\sim eBV_wR_s = 50$ TeV \citep{Morlino_2021}. 

    The winds in star clusters can naturally develop Kolmogorov turbulences \citep{Gallegos_2020}. Assuming this regime for the magnetic field turbulence inside the shocked region, the diffusion coefficient can be estimated as \citep{Morlino_2021}
    \begin{align}
        \nonumber
        D(E,r) \approx \frac{1}{3}r_L(E)v_W\left(\frac{r_L(E)}{L_c}\right)^{-2/3}\left(\frac{r}{R_s}\right)^{1/3} = \\
        = 4.0\cdot10^{26} \textrm{\;cm}^2/\textrm{s}   ~\left(\frac{E}{1 \textrm{\; GeV}}\right)^{1/3} \left(\frac{r}{R_s}\right)^{1/3},
    \end{align}
    with $r_L$ the Larmor radius of the particles, $L_c$ the size of the cluster and $R_s$ the radius of the shock. Outside this region we use $D_0 = 2.0\cdot10^{28} \mathrm{\;cm}^2\,\mathrm{s}^{-1}$ as normalization at 1 GeV to reproduce the interstellar Galactic propagation, but keeping the same scaling with energy. This value is an estimate based on the local cosmic-ray measurements' allowed range \citep[e.g.][for opposite extremes of this range]{Evoli_2019,Thaler_2023}.
    
    The region contains a large number of O- and Wolf-Rayet stars, which have wind velocities between 2000 and 3000 km/s \citep{Prinja_1990}. We then assume a radial collective wind with velocity $v_W =$ 2500 km/s.
    
    MHD simulations have shown that the magnetic field inside star-forming regions varies much depending on the exact position, with drastic changes on scales of $\sim 0.05$ pc and with magnetic field values ranging from a few to hundreds of $\mu$G \citep{Badmaev_2022}. To model such turbulent magnetic field is beyond the scope of this work. We instead consider a constant value of 7 $\mu$G for the whole shocked region. This value is an average over the resulting magnetic field from a MHD simulation of the innermost region of a star cluster \citep{Badmaev_2022,Bykov_2023}.


\section{Gamma-ray observations}
\label{sec:analysis}

     In order to observationally constrain the non-thermal output of our model, we perform a morphological and spectroscopic analysis of NGC 3603 in the GeV photon band. The LAT is a pair-production $\gamma$-ray detector, part of the \textit{Fermi Gamma-ray Space Telescope} mission \citep{LATpaper}. In operation since August 2008, it performs an all-sky survey in the $\gamma$-ray band from 30 MeV to more than 500 GeV. Its point-spread function (PSF) is energy-dependent. It varies from several degrees at the lowest energies to less than $0.1^{\circ}$ at 10 GeV \citep{4FGL}.
     Those values refer to average quality-reconstruction data, hence we will perform an optimized analysis through different cut selections.

    In this work, we perform a binned maximum-likelihood analysis fit \citep{Mattox96} on more than 15 years of LAT data (August 4 2008 to January 18 2024). We select P8R3 \texttt{SOURCE} data \citep[evclass $=128$;][]{Atwood13, Bruel18} between 1 GeV to 1 TeV, with a cut on the zenith angle $z_{\rm max}$ at $105^{\circ}$. NGC~3603 being an established bright source at those energies \citep[4FGL J1115.1-6118;][]{Saha_2020}, we can apply a cut on the direction reconstruction quality, using only the best-quality data quartile (PSF3, evtype $=32$).

    The data is binned in a $10^{\circ}\times10^{\circ}$ grid with $0.05^{\circ}$ spatial bins and $8$ energy bins per decade, while the analysis itself is performed using the \texttt{Fermitools} \citep[v2.2.0;][]{Fermitools} and \texttt{fermipy} \citep[v1.2.0;][]{Wood17}. As a source background model we employ the 4FGL-DR4 catalogue \citep[v33;][]{4FGL, Ballet23}, with `gll\_iem\_v07.fits` and `iso\_P8R3\_SOURCE\_V3\_PSF3\_v1.txt` as the Galactic and isotropic diffuse components, respectively. Energy dispersion correction is applied to all sources except the isotropic diffuse component. In this setup, the detection significance is evaluated considering the test statistic $\rm{TS} = - 2\;\ln L_{\textrm{max},0} /L_{\textrm{max},1}$, where $L_{\textrm{max},0}$ is the likelihood value for the null hypothesis and $L_{\textrm{max},1}$ the likelihood for the tested model. A $\rm{TS}=25$ roughly corresponds to a detection at $5\sigma$ and $4\sigma$, for 1 and 4 source parameters, respectively.
    
    Our analysis is performed in three steps: (1) we optimize the region of interest (ROI) through an iterative process, fitting the normalization of all sources in descending order of their predicted number of counts ($N_{\rm pred}$) down to $N_{\rm pred}=1$ (\texttt{fermipy.optimize}). Then, (2) we free the normalization of all sources within $6^{\circ}$ of the ROI centre, and fit them with \texttt{NEWMINUIT}. Finally, (3) we search for new sources ($>4\sigma$, $>0.3^{\circ}$ from known sources), re-localize the nearby extended source 4FGL J1109.4-6115e ($r= 1.27^{\circ}$, at $0.69^{\circ}$ from NGC 3603), and repeat step (2) but freeing all spectral parameters for those sources previously detected above $5\sigma$. These steps result in a good data-model agreement as seen in the p-value statistic \citep[PS; ][]{Bruel21} residuals of our ROI (Appendix \ref{app:residuals}), and the addition of seven putative sources, all far away from NGC 3603 ($>3.4^{\circ}$; Appendix \ref{app:sources}). The centroid of 4FGL J1109.4-6115e is displaced to ($l$, $b$) = ($290.634\pm0.082$, $-0.773\pm0.067$), with an offset of $0.35^{\circ}$ with respect to its catalog position and $1.0^{\circ}$ from NGC 3603. 4FGL J1115.1-6118 is detected with $\rm{TS} = 163.5$, $\Gamma= 2.54\pm0.13$, an integrated photon flux above 1 GeV of $(1.94 \pm 0.22) \times 10^{-9}$ photons cm$^{-2}$ s$^{-1}$.

    \begin{figure}
        \centering
        \includegraphics[width=\hsize]{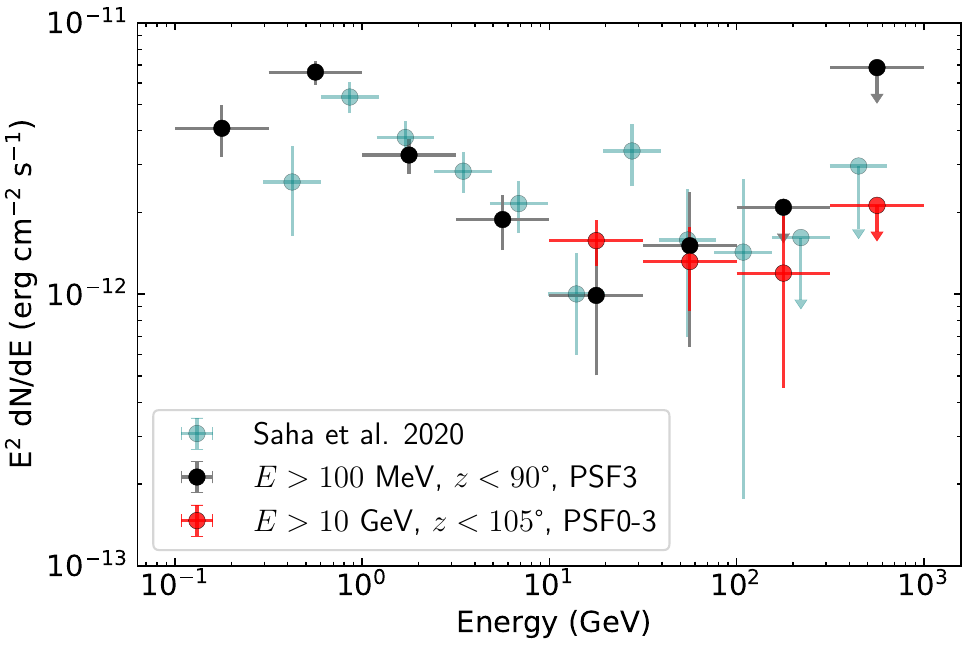}
        \caption{Resulting SED for the datasets above 100 MeV and 10 GeV, respectively, and compared with the results from \cite{Saha_2020}. Upper limit from \textit{Fermi}-LAT at 95\% confidence level are reported for energy bins detected at less than $2\sigma$.}
        \label{fig:fermi_sed}
    \end{figure}

    The previous description provides an optimal ROI background at the energies where LAT performs best, yet it is insufficient to achieve either the spectral or morphological objectives of the present analysis -- this is, provide the required information to identify the parent particle population and resolve the source extension below $0.1^{\circ}$. To this end we employ two additional analyses: setting the low energy thresholds at 100 MeV and 10 GeV, respectively. While the latter can make use of the same setup, the former requires an update of the ROI to $20^{\circ}\times20^{\circ}$ (with spatial bins of $0.1^{\circ}$) and a reduced maximum zenith angle of $90^{\circ}$ to avoid Earth-limb contamination. In both cases, we employ the resulting model from the 1 GeV threshold analysis as our initial background, and then perform again steps (1), (2), and searched for new sources, while setting the spectral parameters of 4FGL J1115.1-6118 free (see Figure~\ref{fig:fermi_sed} and Table~\ref{tab:lat}). Again, we validate the fit through the distribution of PS residuals.

    The latter analysis setup, i.e. above 10 GeV ($\rm{TS}= 28.3$), provides an upper limit of 0.096$^\circ$ for the extension of 4FGL J1115.1-6118 at a 95\% confidence level (TS$_{\rm ext} = 2.3$). This result could be highly dependent on the underlying diffuse model \citep{Saha_2020}, therefore we test for the extension also employing an older version of it (`gll\_iem\_v06.fits`). We find the lack of extension to be robust under the PSF3 selection (TS$_{\rm ext} = 0.6$), with the 95\% upper limit set at 0.084$^\circ$. We note, however, that the detection is less significant, as 4FGL J1115.1-6118's TS goes down to 19.3. 
    
    In addition, it is relevant to assess the possible presence of a distinct component in the same energy range, already discussed in \cite{Saha_2020}, as well as its spectral shape. Therefore we repeat the analysis above this energy threshold, but now without the constraint on the quality reconstruction (i.e. a joint analysis using event types 4, 8, 16, and 32; PSF0--3). This increases in about 4 times the number of events, which allows a better assessment of the spectral characteristics, with a source detection at $\sim 10\sigma$. Despite the larger PSF, the increased statistics facilitate as well a better localization ($\sim 50\%$ improvement) of the centroid of the gamma-ray source ($l$,$b$) = ($291.624 \pm   0.012$, $-0.553 \pm0.013$). That is consistent with the 4FGL-DR4 localization and the center of NGC 3603. Such a selection leads to a worse extension constraint due to a larger preference for extension (TS$_{\rm ext} = 11.1$, 95\% upper limit at 0.106$^\circ$). Finally, we test the presence of curvature assuming a power-law and log-parabola, finding no preference for curvature ($\Delta\rm{TS}=2.2$), but well represented by a hard spectrum ($\Gamma =2.0\pm0.1$) -- which does not differ at more than $4\sigma$ from that of a single power-law above 1 GeV.

\begin{table*}[h]
\caption{\textit{Fermi}-LAT analysis results for NGC 3603.}             
\label{table:etacar}      
\centering                          
\begin{tabular}{c c c c}        
\hline\hline      
Parameter & 100 MeV -- 1 TeV & 1 GeV -- 1 TeV & 10 GeV -- 1 TeV (PSF0--3)\\  
\hline     
TS & 230.1 & 163.5 & 28.3 (99.6)  \\
$\Gamma$ & $2.30\pm 0.06$ & $2.54 \pm 0.13$ & $1.84 \pm 0.34$ ($2.01\pm0.05$)  \\
Energy Flux ($10^{-12}$ erg cm$^{-2}$ s$^{-1}$) & $23.6 \pm 1.5$ & $8.69 \pm 1.15$ & $5.46\pm 3.04$ ($6.51\pm 1.05$) \\
Photon Flux ($10^{-9}$ ph cm$^{-2}$ s$^{-1}$) & $36.3 \pm 5.7$ & $1.94 \pm 0.22$& $0.06\pm 0.02$ ($0.09\pm 0.01$)\\

\hline  
\end{tabular}
\label{tab:lat}
\end{table*}

\section{Results}
\label{sec:results} 
    
    It is mainly the gamma-ray spectrum that is used to constrain the free parameters in our model. We test the injection spectral index $\alpha$ in a range $\alpha \in \left[2.2,2.8 \right]$ in steps of 0.05, performing a chi-square fit of the gamma-ray spectrum to find the best acceleration efficiency for every single slope. The final model is the one with the slope resulting in the best chi-square result after the efficiency fitting. To calculate the gamma-ray fluxes, we take into account pion production, initiated by CR protons, as well as IC scattering and non-thermal bremsstrahlung, by CR leptons. For the former, we use of the cross sections from \cite{Kamae_2006}, already included in PICARD, for computing the proton-proton interactions.

    In addition, we also considered the effect of $\gamma\gamma$-pair production on the resulting gamma-ray emission. Despite the high number of UV photons in comparison to the typical ISRF models in the Galaxy (see Figure~\ref{fig:stars}), the size of the star-forming region is too small for the absorption to make an important contribution given the $\gamma\gamma$-cross-section. The flux is decreased by only $\sim$0.8\% at the energy of maximum effect ($\sim$100 GeV).

\subsection{Hadronic scenario}

    \begin{figure}
        \centering
        \includegraphics[width=\hsize]{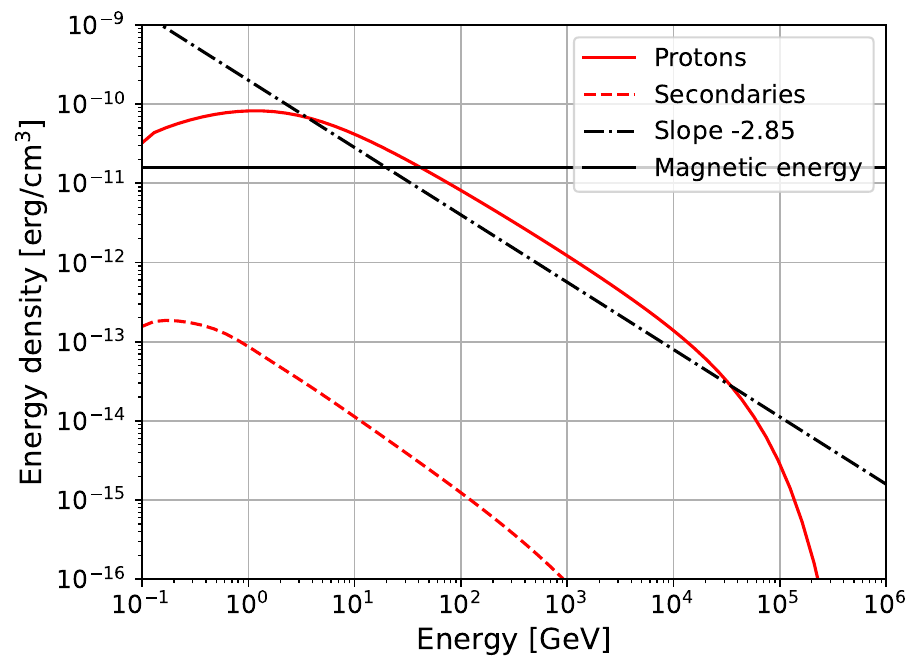}
        \includegraphics[width=\hsize]{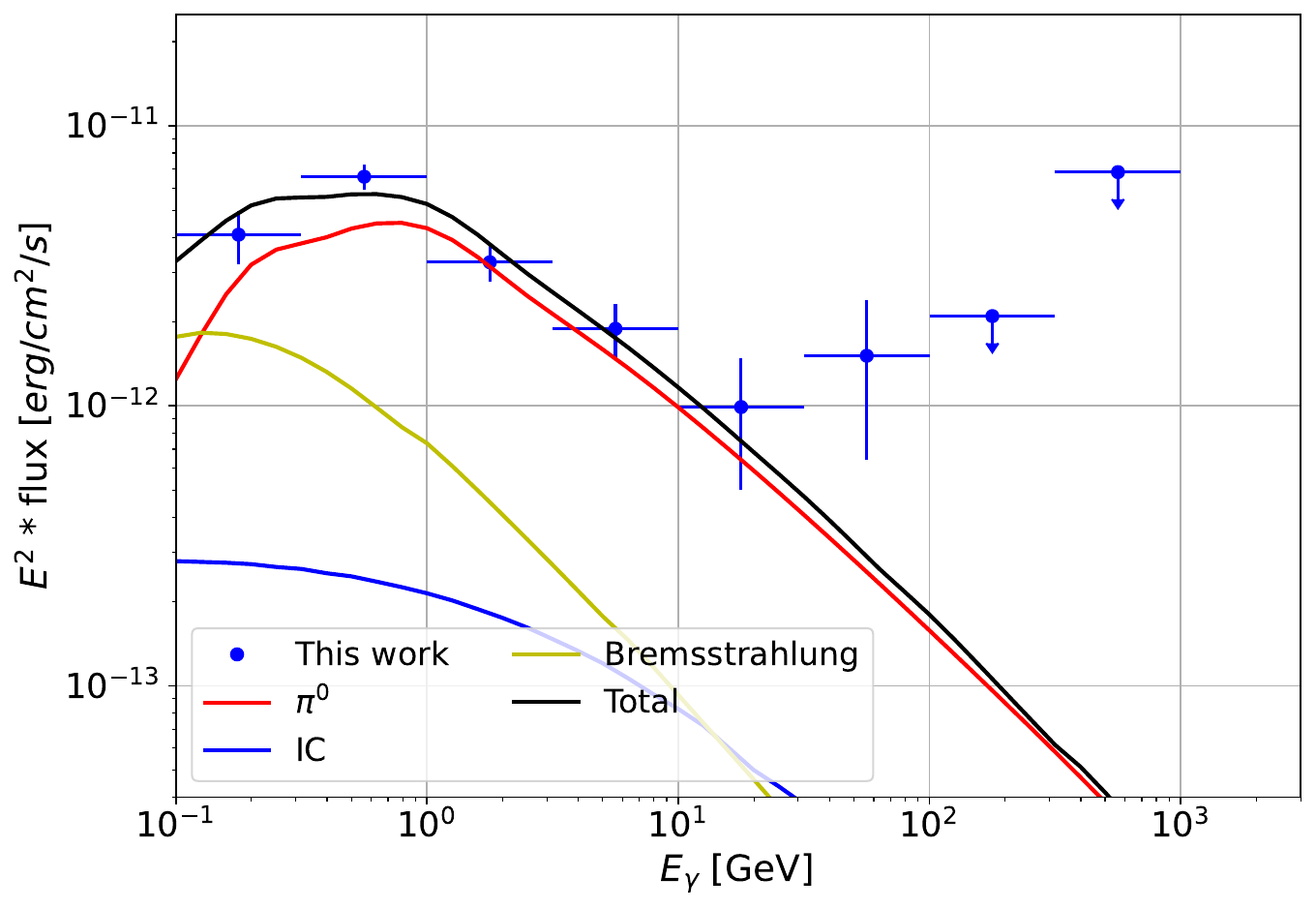}
        \caption{Cosmic-ray (top) and gamma-ray (bottom) spectrum in the hadronic scenario. Secondaries are electrons and positrons created by CR protons in pp-interactions. The gray band shows the H.E.S.S. sensitivity for 50 h of observations \citep{Holler_2015}.}
        \label{fig:hadr_spec}
    \end{figure}

    We show the resulting cosmic-ray spectra of the hadronic scenario in the upper panel of Figure~\ref{fig:hadr_spec} and their associated spectral energy distribution in the lower one. In this case, the best-fit scenario results in an injection index of $\alpha_{had}$ = 2.6. However, this result faces the problem of an unusually high acceleration efficiency, nearly 30\%, that is needed. 

    \begin{figure}
        \centering
        \includegraphics[width=\hsize]{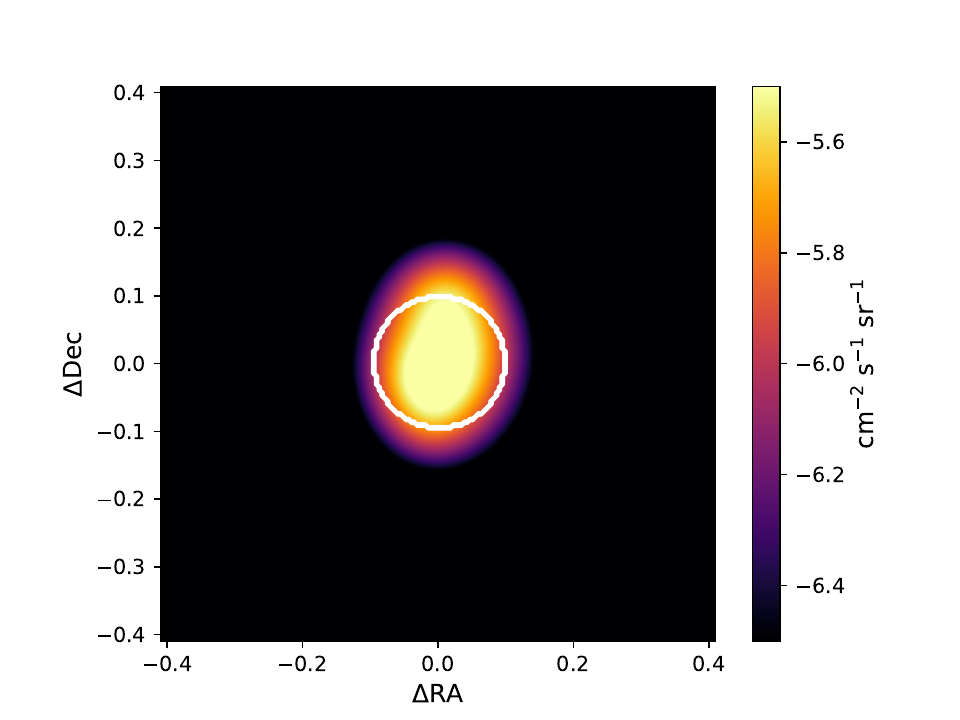}
        \caption{Resulting integrated gamma-ray map above 10 GeV in the hadronic scenario. The white circle shows the extension upper limit from \textit{Fermi}-LAT at 95\% confidence level.}
        \label{fig:hadr_ext}
    \end{figure}

    In order to compare the $\gamma$-ray extension of our model with the observational result, we need to account for the LAT PSF. Therefore, we smooth our resulting $\gamma$-ray maps using a Gaussian kernel with the PSF of the instrument \citep{Fermi_performance}. The resulting emission map is shown in Figure~\ref{fig:hadr_ext}. The extension of the $\gamma$-ray emission originating from our model is calculated as the $1\sigma$ containment radius, to be comparable with the data analysis. It gives an extension of 0.098$^\circ$ for energies above 10 GeV. This result is in mild tension with the data upper limit, especially considering our limited knowledge on the Galactic diffuse model. Our predicted $\gamma$-ray extension is shown in Table~\ref{tab:extension}.
    We also studied the effect of varying the diffusion coefficient outside the shocked region, since we were using an estimated value in the middle of the allowed range by local measurements (Section~\ref{sec:model}). However, using $D_0$ = 10$^{28}$ cm$^2$s$^{-1}$, i.e. half of our default normalization outside the shocked region, does not affect the $\gamma$-ray extension visibly.

\subsection{Leptonic scenario}

    \begin{figure}
        \centering
        \includegraphics[width=\hsize]{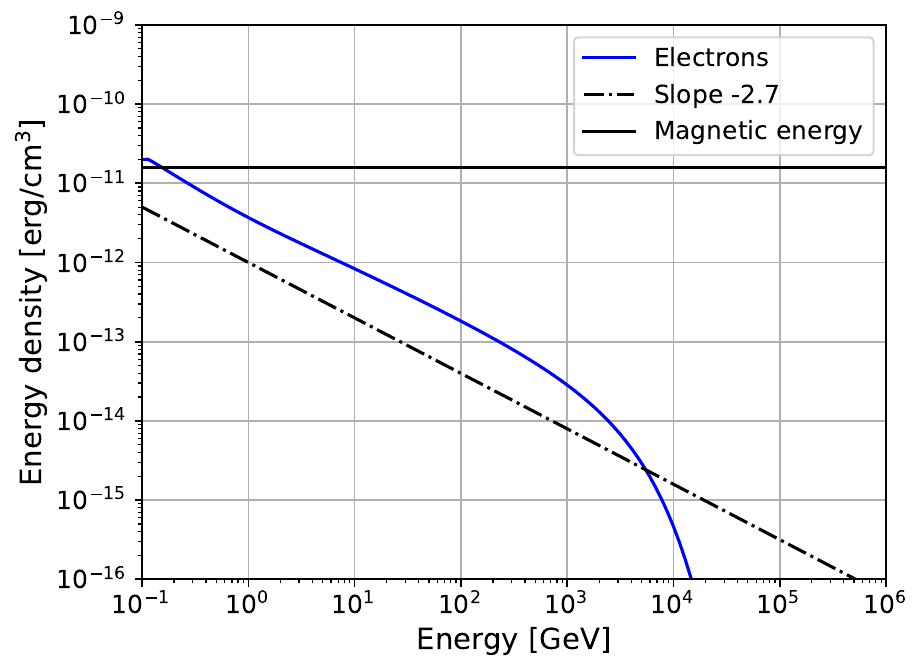}
        \includegraphics[width=\hsize]{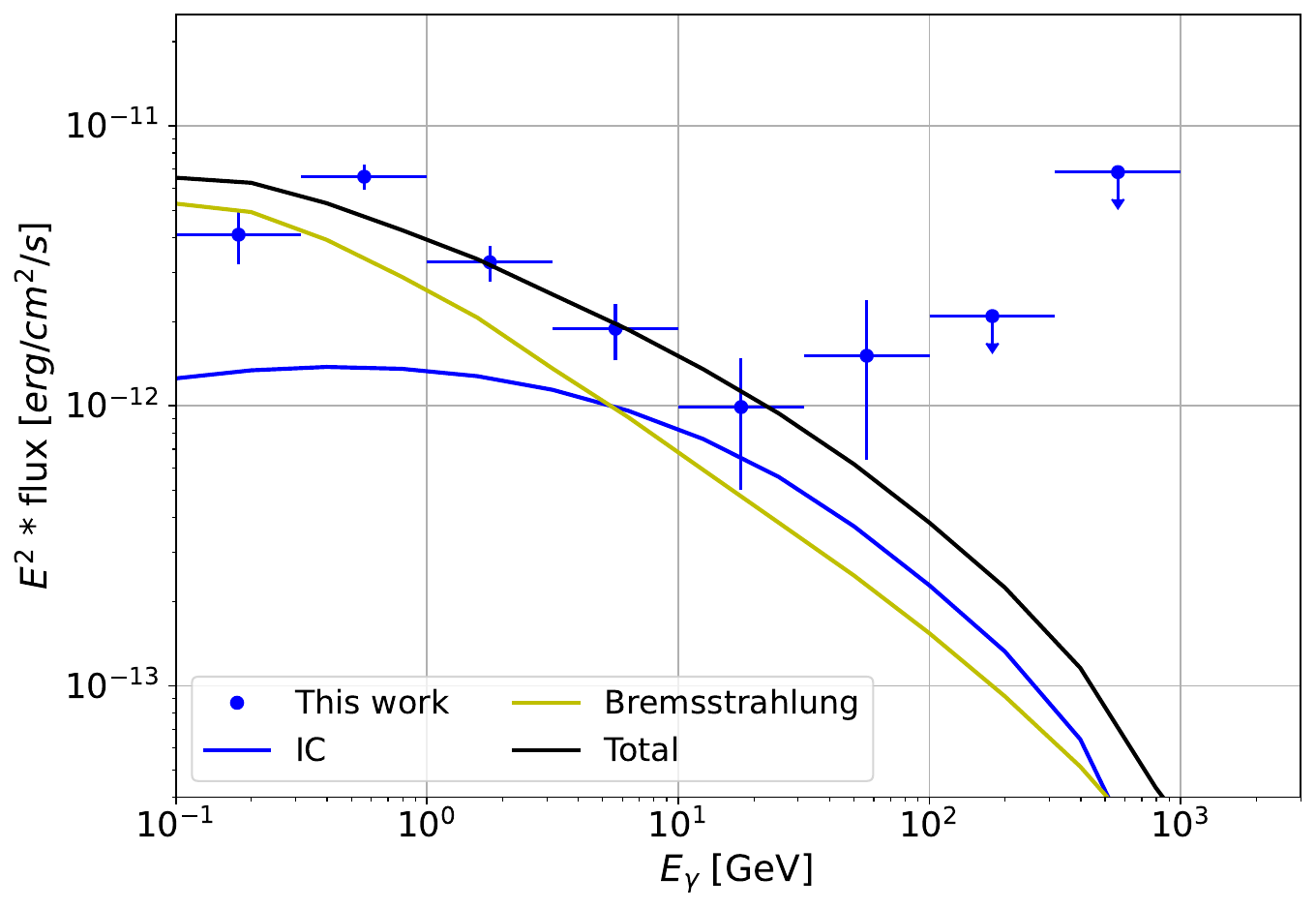}
        \caption{Same as Figure~\ref{fig:hadr_spec} but for the leptonic case.}
        \label{fig:lept_specs}
    \end{figure}

    The results of this scenario are shown in Figure~\ref{fig:lept_specs}. In this case, the best-fit injection particle spectral index is $\alpha_{lep}$ = 2.5, while the acceleration efficiency is found to be around 0.4\% -- much less energy is needed to reach the measured flux. As one can see in the lower panel of Figure~\ref{fig:lept_specs}, the model does not well reproduce the observed gamma-ray spectrum. The data presents a bump-like feature at low energies, which is completely missed in this scenario. In this case, however, the resulting extension is 0.086$^\circ$, consistent with the data upper limit

\subsection{Hybrid scenario}

    \begin{figure}
        \centering
        \includegraphics[width=\hsize]{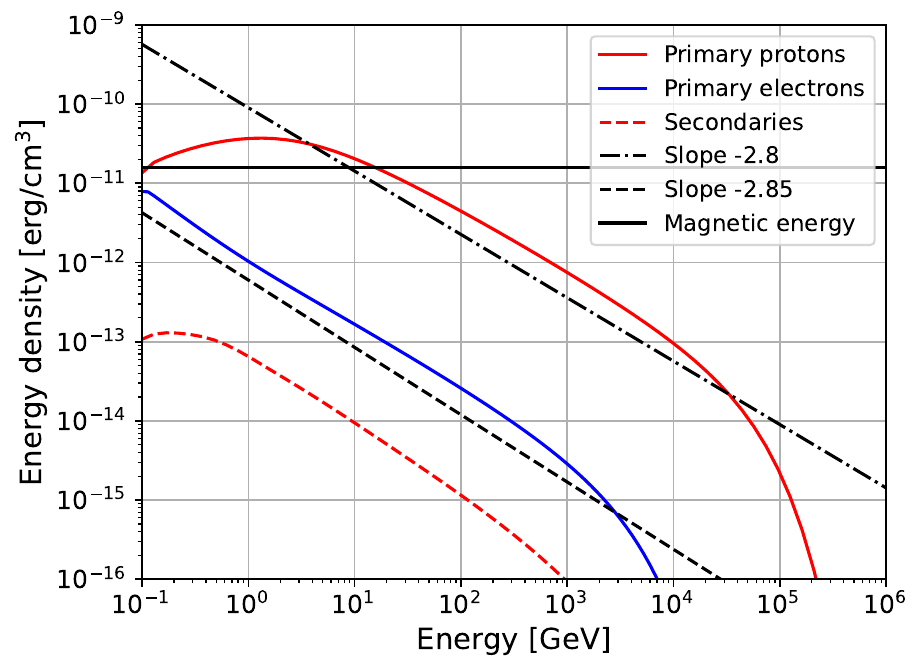}
        \includegraphics[width=\hsize]{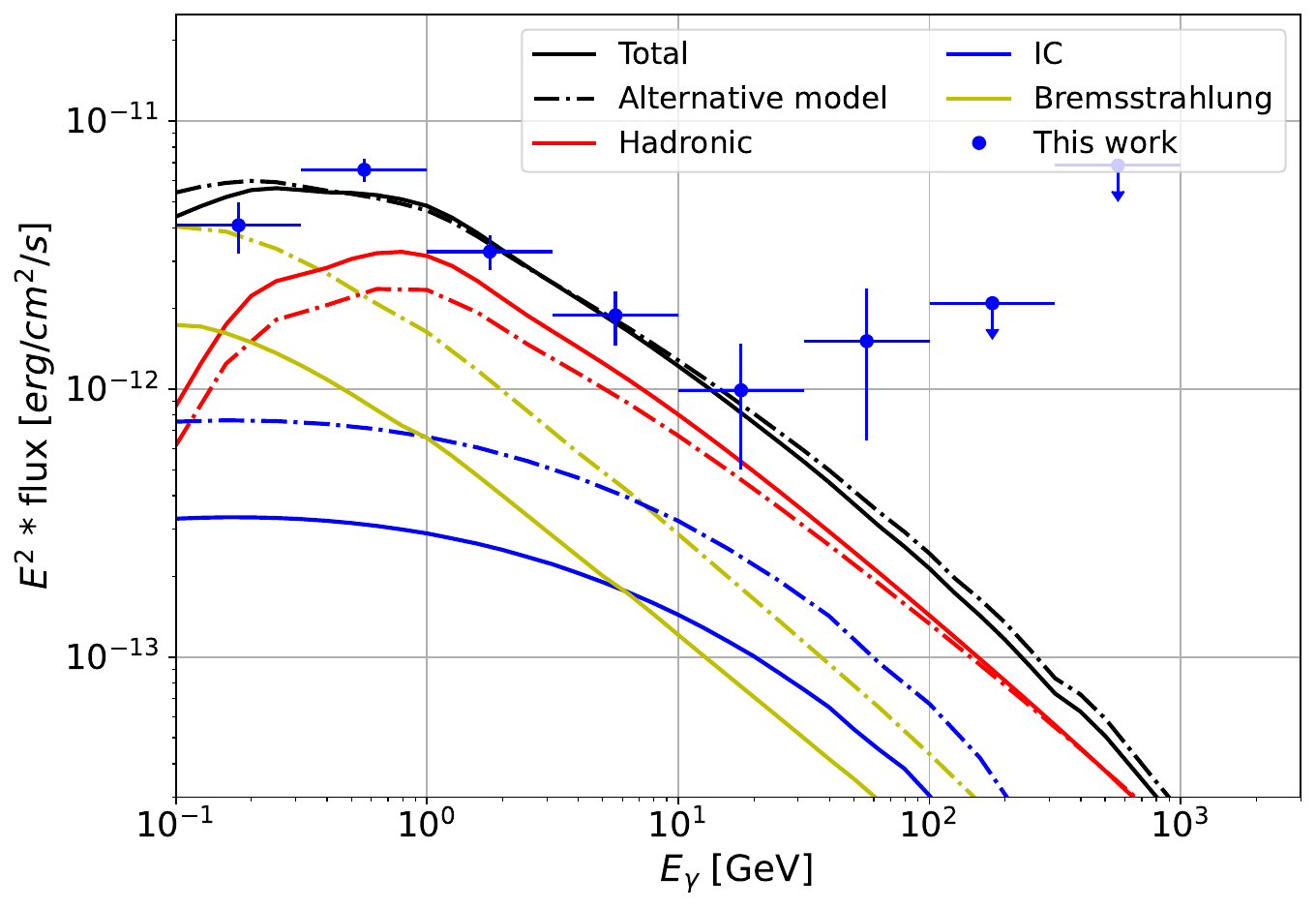}
        \caption{Same as Figure~\ref{fig:hadr_spec} but for the hybrid scenario. The dashed lines in the gamma-ray spectrum show an alternative option to the best-fitting result.}
        \label{fig:hyb_specs}
    \end{figure}

    The fitting results in the hybrid scenario can be seen in Figure~\ref{fig:hyb_specs}. The best-fit result to the gamma-ray spectrum leads to an injection of protons with an index $\alpha_{hyb,p}$ = 2.55, electrons with $\alpha_{hyb,e}$ = 2.65, with an acceleration efficiency around 20\% and a $n_p /n_e$ injected ratio of 200. In this case, the gamma-ray spectrum is still well reproduced, and the extension of the emission is 0.096$^\circ$. On the other hand, if we slightly alter the parameter values, the extension can be lowered with still a decent fit to the gamma-ray spectrum. For example, when using $\alpha_{hyb,p}$ = 2.5, $\alpha_{hyb,e}$ = 2.65, the efficiency needed is reduced to about 10\% and the $n_p/n_e$ ratio gives 80. The resulting gamma-ray spectrum is shown as the dash-dotted line in the lower panel of Figure~\ref{fig:hyb_specs} and the extension is 0.093$^\circ$, consistent with the data.

    \begin{table*}
        \caption{Extension of the $\gamma$-ray source in the data and the different scenarios above 10 GeV. Introducing the simulated $\gamma$-ray maps and spectra into our likelihood analysis framework, $\Delta \rm{TS}_{\Gamma=2.54}$ highlights the preference for the different scenarios. Each case is compared with the result from a point-like source with $\Gamma=2.54$.}
        \centering
        \begin{tabular}{|ccccc|}
        \hline
          & $\alpha$ & $\eta$ [\%] & Extension [deg] & $\Delta \rm{TS}_{\Gamma=2.54}$ \\ \hline
        Data & - & - & $< 0.096$ (at 95\%; $\rm{TS}_{\rm{ext}}=2.3$ ) & - \\ 
        Hadronic & 2.6 & 26 & 0.098 & $-4.6$\\ 
        Leptonic & 2.5 & 0.38 & 0.086 & $-2.7$\\ 
        Hybrid & 2.55 (p), 2.65 (e$^-$) & 19 & 0.096 & $-4.4$\\ 
        Alt. hybrid & 2.5 (p), 2.65 (e$^-$) & 13 & 0.093 & $-4.2$\\ \hline
        \end{tabular}
        \label{tab:extension}
    \end{table*}

\subsection{Other wavelengths and observables}
    
    \begin{figure}
        \centering
        \includegraphics[width=\hsize]{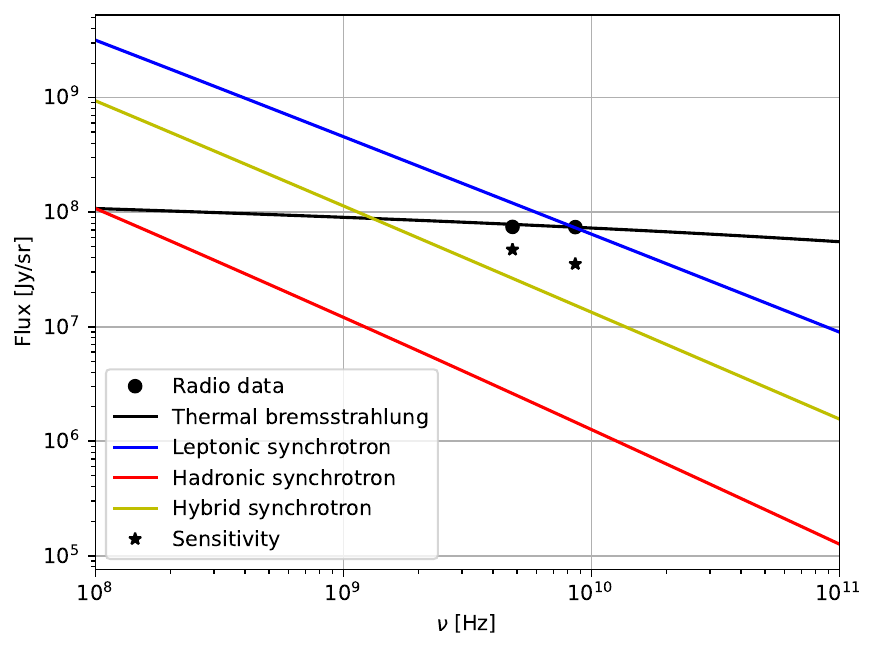}
        \caption{Radio flux from NGC 3603. The black solid line shows a thermal bremsstrahlung emission coming from an electron population with $10^4$ K and black dots are the measured data from \cite{Mucke_2002}. The star markers represent the sensitivity limit of the ATCA observations.}
        \label{fig:radio}
    \end{figure}

    \begin{figure}
        \centering
        \includegraphics[width=\hsize]{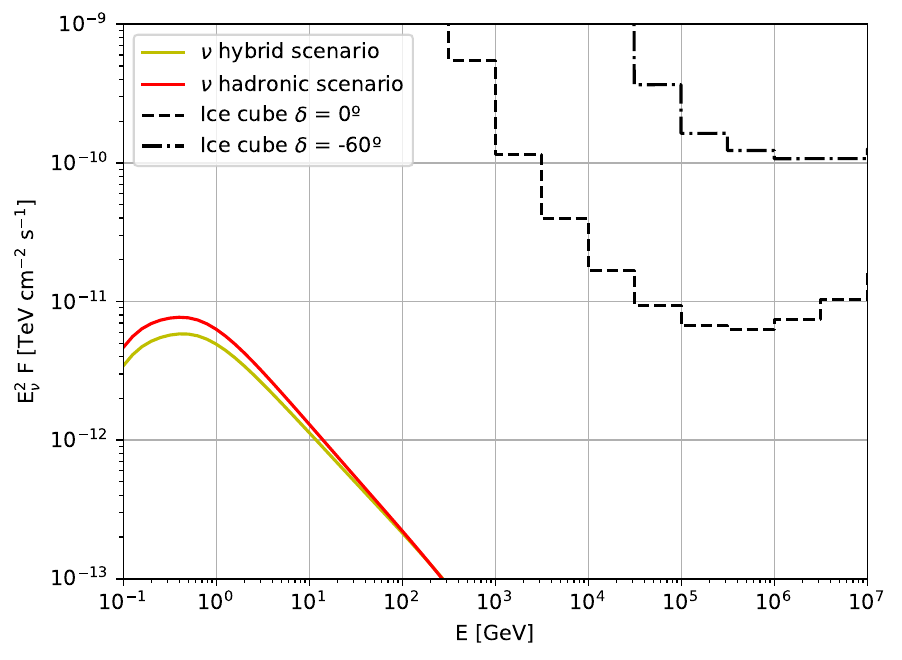}
        \caption{Neutrino flux from NGC 3603. The black dashed and dash-dotted lines represent the IceCube experiment sensitivity for different declinations \citep{Aartsen_2017}. }
        \label{fig:neutrino}
    \end{figure}

    In order to further constrain the contributions of each particle population to the total gamma-ray emission, we also consider other wavelengths and observables, such as radio and X-ray radiation and neutrino emission. NGC 3603 was already studied in the radio domain by \cite{Mucke_2002} with the Australia Telescope Compact Array (ATCA) to search for possible non-thermal signals. However, their results were inconclusive, neither a cosmic-ray nor a thermal component from cold dust could explain the measured data alone.
    
    In contrast to the data, which resembles a quasi-point-like source ($\sim$ 1 arcsec) at the center of the star cluster, the results of our simulations are much more extended ($\sim$ 30 arcsec). With this incompatibility in size, our models would need to be below the sensitivity of the observations in order to be acceptable. We show the synchrotron surface brightness in Figure~\ref{fig:radio}, together with the measured data and the sensitivity of the instrument, given by the r.m.s of the observations' noise. The results clearly disfavour a leptonic-only scenario, whose synchrotron emission is above the sensitivity of the ATCA observations. The hybrid and hadronic scenarios are compatible with the radio data, since they are both below the sensitivity. As already suggested in \cite{Nurnberger_2003}, the data seems to be better reproduced with a thermal origin. This hypothesis is also shown in the figure with a model for the thermal bremsstrahlung from a population of electrons with temperature $10^4$ K. If instead we consider X-ray wavelengths, our resulting electron population would produce a non-thermal flux 4 orders of magnitude below the diffuse thermal component measured by \textit{Chandra} in the 0.5-7 keV energy range \citep{Townsley_2011}. 
    
    High-energy neutrinos are produced in cosmic-ray proton interactions, therefore they represent a good tracer of the hadronic origin of gamma rays. We calculate the neutrino production for our scenarios containing relativistic protons using the Kamae cross-sections \citep{Kamae_2006}. Our results are shown in Figure~\ref{fig:neutrino}. All are far below the sensitivity of the IceCube experiment \citep{Aartsen_2017} for this sky region, whose declination is $\delta \sim -60^\circ$. Other instruments, such as ANTARES or KM3Net are also unlikely to detect it. This is mainly because of the low energy of the neutrinos, whose peak is at $\sim$1 GeV, while the instruments are optimized for higher energies.


\section{Discussion}
\label{sec:discussion}
    
    This approach to model the extension in the hadronic scenario is robust, since it does not depend on the propagation setup. Due to the small size of the region, even when reducing the diffusion coefficient, the particles are able to reach the outer molecular clouds, where the gamma rays are produced. Therefore, the results are mainly determined by the angular size of the gas in the observations. In contrast to protons, electrons only propagate a limited distance, which will appear larger if the source is closer. Hence, the resulting leptonic extension is more subject to the uncertainty in the distance. Recent measurements of the NGC 3603 distance result in $7.2\pm0.1$ kpc \citep{Drew_2019}, which already makes our calculations conservative.

    On one hand, the dense molecular clouds surrounding NGC 3603 may serve as target material for the production of gamma rays. Hence, in the hadronic scenario, the more surrounding material, the higher the produced gamma-ray flux. This scenario can provide a good fit of the spectrum, but the extension of the simulations is in tension with the data upper limits. It can be solved introducing some amount of primary electrons. In the alternative hybrid scenario we presented, 45\% of the total gamma-ray energy has a primary leptonic origin, decreasing the extension to 0.093$^\circ$ for energies above 10 GeV. For a more in depth comparison, we introduce the simulated map and spectrum for each scenario in our gamma-ray likelihood analysis above 10 GeV. We compare the results with those of a point-like source with $\Gamma=2.54$ (i.e. the photon index above 1 GeV). The background remains fixed and therefore the normalization is the only degree of freedom in each scenario. We found the point-like source to be preferred. However, such overall morphological preference for a point-like source is solely at the level of $\Delta\rm{TS}=3$--$5$ (Table \ref{tab:extension}).

    On the other hand, the dense clouds make it more difficult for the bubble (and the shock) to form and grow. The necessary efficiency in the hadronic scenario is rather high compared to the standard 10\% value. However, there are studies indicating that star-forming regions could reach efficiencies of 30\% in the early stages of the bubble \citep{Bykov_1999}. This might be the case for NGC 3603, with an age of only 1-2 Myr \citep{Melena_2008}. The index of the measured gamma-ray spectrum ($\Gamma = 2.54$) might be a consequence of this young age and the dense surrounding molecular clouds. The shock might not be completely formed, leading to softer indices compared to older star-forming regions \citep[e.g. Westerlund 1;][]{Aharonian_2022}.

    Considering alternative source scenarios for the gamma-ray origin, this system is too young to contain supernova remnants or pulsar wind nebulae. Furthermore, the closest source in the ATNF catalog\footnote{\url{https://www.atnf.csiro.au/research/pulsar/psrcat/}} \citep{pulsar_catalog} is PSR J1114-6100, which has no association listed in the catalog. Nevertheless, it is located at a distance of $\sim$0.28$^\circ$, outside the detected gamma-ray source. The only other possible scenario, so far disregarded, would be that of a single star or stellar binary being primarily responsible for the radio and gamma-ray emission from this region. The three WR systems hosted by NGC 3603 are responsible for most of the cluster's wind kinetic power \citep{Crowther98}. Most likely, some of them are in colliding-wind binary (CWB) systems \citep{Moffat02}, as at least NGC~3603--A1 (WR 43a) and NGC 3603--C (WR 43c) have known orbital periods \citep[3.77 and 8.89 days, respectively;][]{Schnurr08}. A priori, the most suitable candidate appears to be NGC~3603--A1, coincident with the radio emission at the core of the stellar cluster \citep{Moffat02} and within the 95\% localization of 4FGL~J1115.1-6118's centroid. This is a particularly massive system (total mass likely larger than $200$~M$_{\odot}$), with a wind kinetic power of $\sim 1.5\times10^{38}$ erg/s, and very similar to the well-studied WR 20a. The appeal of this scenario is fostered by the previous detection of two CWBs by \textit{Fermi}-LAT \citep[$\eta$~Car and $\gamma^2$ Vel; ][]{Abdo10, Marti-Devesa20}, both without strong radio synchrotron emission. Furthermore, $\eta$~Car has an additional hard spectrum beyond 10~GeV \citep{Bednarek11}. If 4FGL~J1115.1-6118 would instead be a CWB, this scenario would be consistent with the point-like nature of the gamma-ray signal and the observed photon index. However, employing photon weights \citep{Kerr19} we do not find a significant periodical modulation in a blind search, nor phase-folding the gamma-ray data with the orbital periods of NGC~3603--A1 or NGC 3603--C. Dedicated simulations of the binary system are necessary to further explore this hypothesis.


\section{Conclusions}
\label{sec:conclusion}

    With the long standing SNR scenario problems to explain all the observable cosmic-ray data, star-forming regions have gained a lot of interest as cosmic-ray sources in the last years. We have studied the case of NGC 3603, a young and compact star-forming region, ideal for the acceleration of high-energy particles. We have performed CR injection and propagation in order to investigate the origin of the gamma-ray emission \citep{Yang_2017, Saha_2020}. In order to do it, we built a realistic 3D model of its radiation field and gas distributions to have a robust result of the extension of the gamma-ray source. 

    Some works have already shown the difficulties for star-forming regions to be PeVatron sources \citep{Morlino_2021, Vieu_2022b} due to their small size. In the case of NGC 3603, even though it is considered a very powerful star cluster, with $L_W \approx 3.2\cdot10^{38} ~\mathrm{erg}~\mathrm{s}^{-1}$, its young age and the dense molecular clouds surrounding it have likely prevented the bubble to grow more. Considering also its distance, NGC 3603 will be complicated to detect by TeV facilities such as H.E.S.S. assuming a cluster origin for the gamma-ray signal, as seen in e.g. Figure~\ref{fig:hadr_spec}.

    Coming back to the main question of this work, the pure hadronic scenario presents a light tension with the data upper limit on the extension of the gamma-ray source. Although this tension is only mild considering the analysis uncertainties, it can be avoided by introducing primary electrons. Primary electrons on their own also have difficulties explaining the multiwavelength data due to conflicts with the radio observations. Therefore, NGC 3603 high energy emissions more likely originate from a mixture of cosmic-ray electrons and protons, with similar importance in order to accommodate the source extension.
    


\begin{acknowledgements}
      The Fermi LAT Collaboration acknowledges generous ongoing support from a number of agencies and institutes that have supported both the development and the operation of the LAT as well as scientific data analysis. These include the National Aeronautics and Space Administration and the Department of Energy in the United States, the Commissariat \`{a} l'Energie Atomique and the Centre National de la Recherche Scientifique / Institut National de Physique Nucl\'{e}aire et de Physique des Particules in France, the Agenzia Spaziale Italiana and the Istituto Nazionale di Fisica Nucleare in Italy, the Ministry of Education, Culture, Sports, Science and Technology (MEXT), High Energy Accelerator Research Organization (KEK) and Japan Aerospace Exploration Agency (JAXA) in Japan, and the K. A. Wallenberg Foundation, the Swedish Research Council and the Swedish National Space Board in Sweden. Additional support for science analysis during the operations phase from the following agencies is also gratefully acknowledged: the Istituto Nazionale di Astrofisica in Italy and the Centre National d'Etudes Spatiales in France. This work performed in part under DOE Contract DE-AC02-76SF00515. 
\end{acknowledgements}

\bibliographystyle{aa} 
\bibliography{bibliography.bib} 

\onecolumn
\begin{appendix}
\section{Data-model validation of the region of interest}\label{app:residuals}

\begin{figure*}[h]
        \centering
        \includegraphics[width=0.49\hsize]{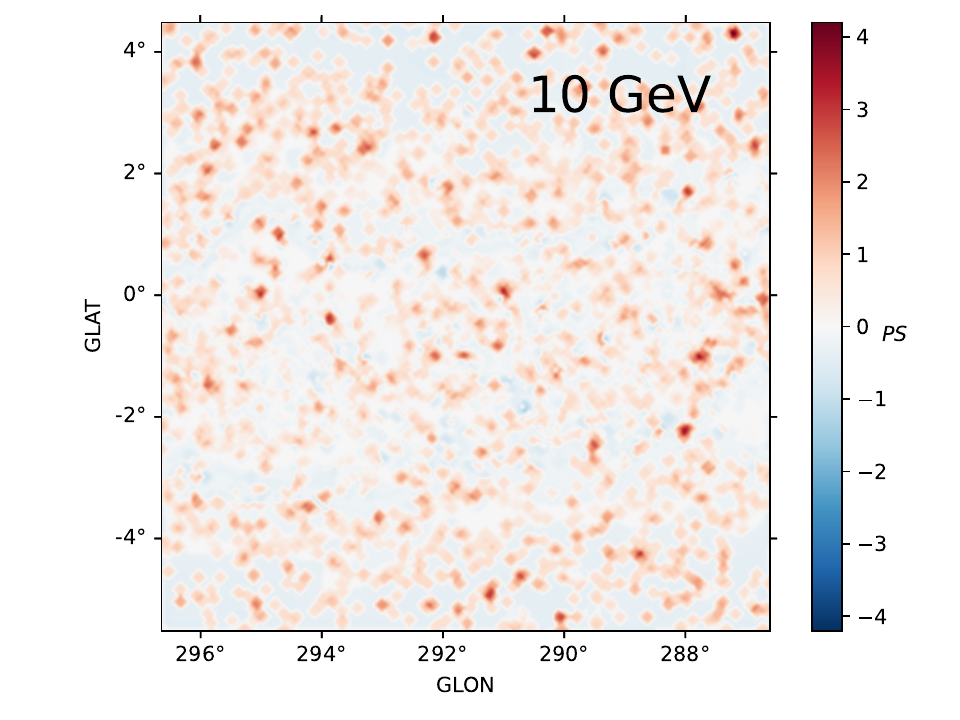}
        \includegraphics[width=0.49\hsize]{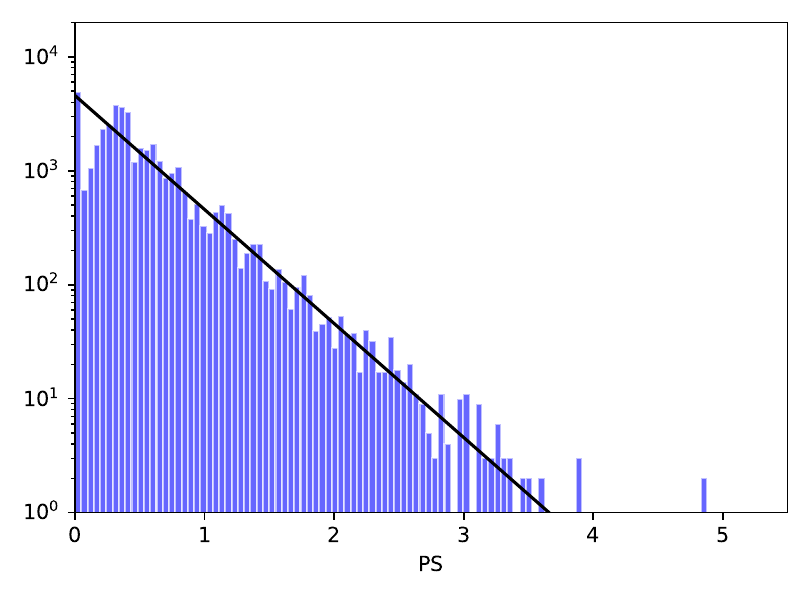}
        \includegraphics[width=0.49\hsize]{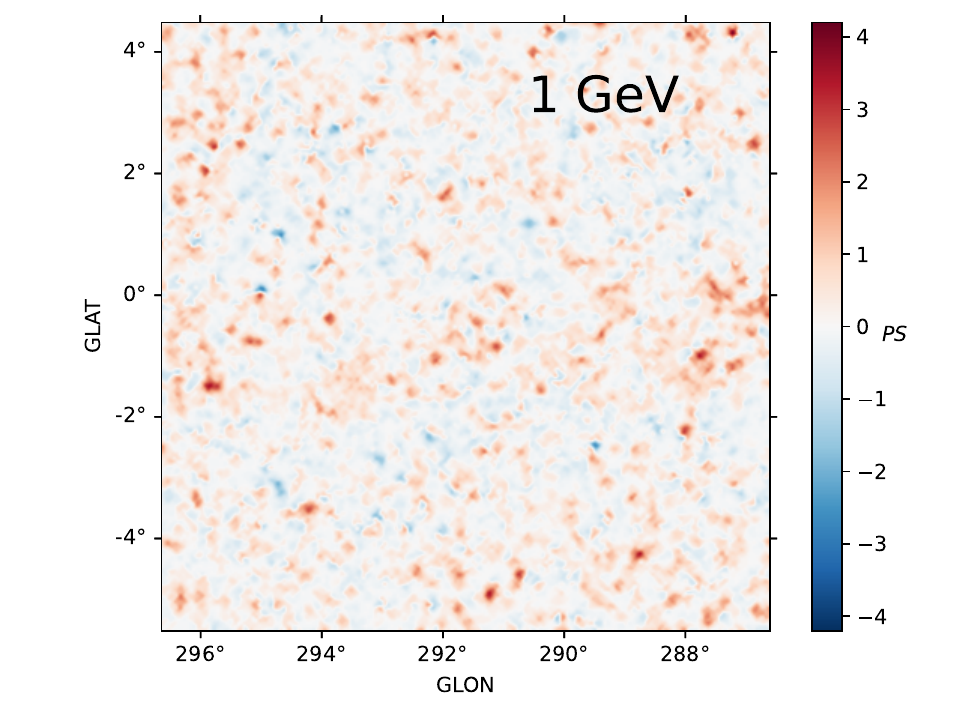}
        \includegraphics[width=0.49\hsize]{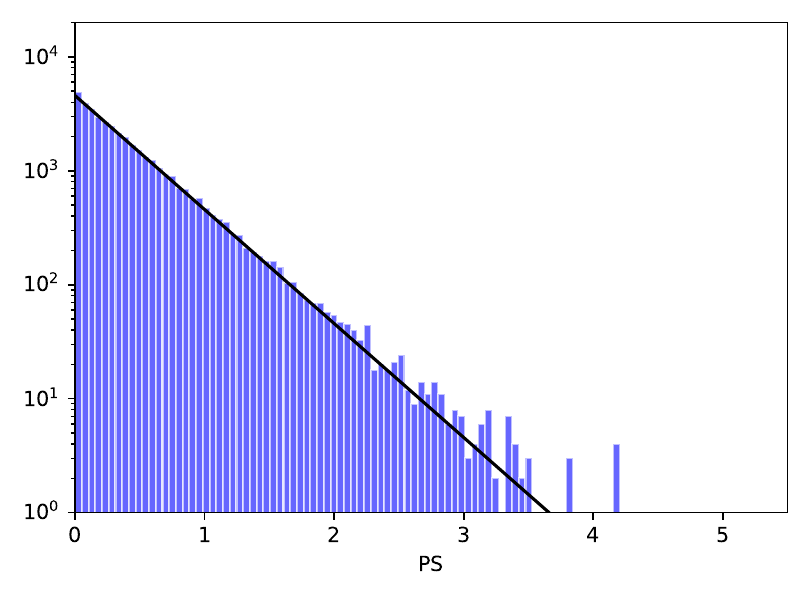}
        \includegraphics[width=0.49\hsize]{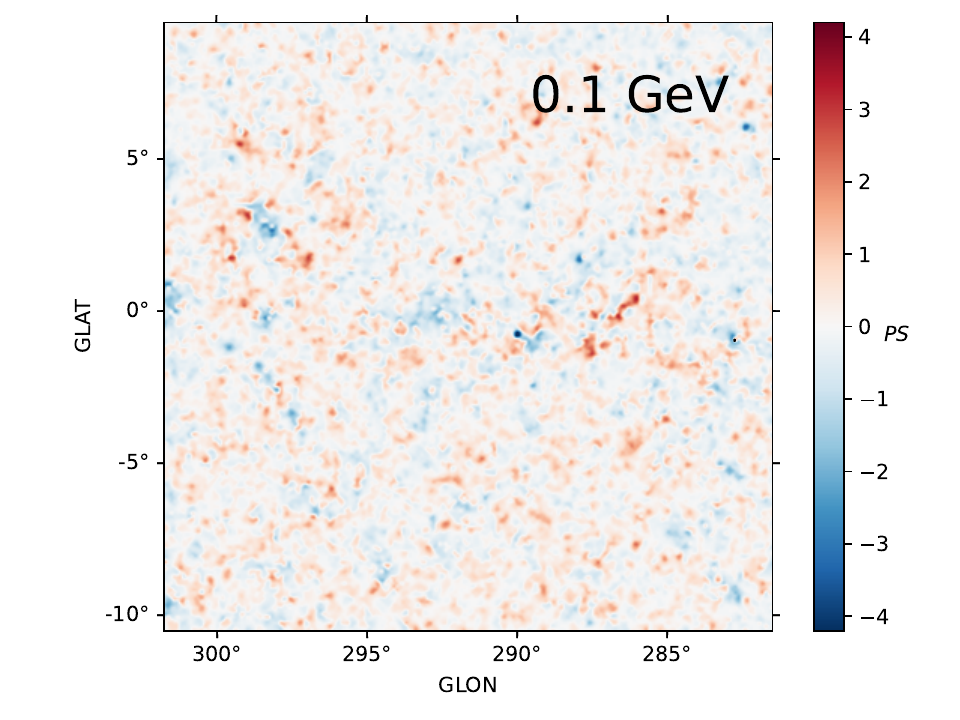}
        \includegraphics[width=0.49\hsize]{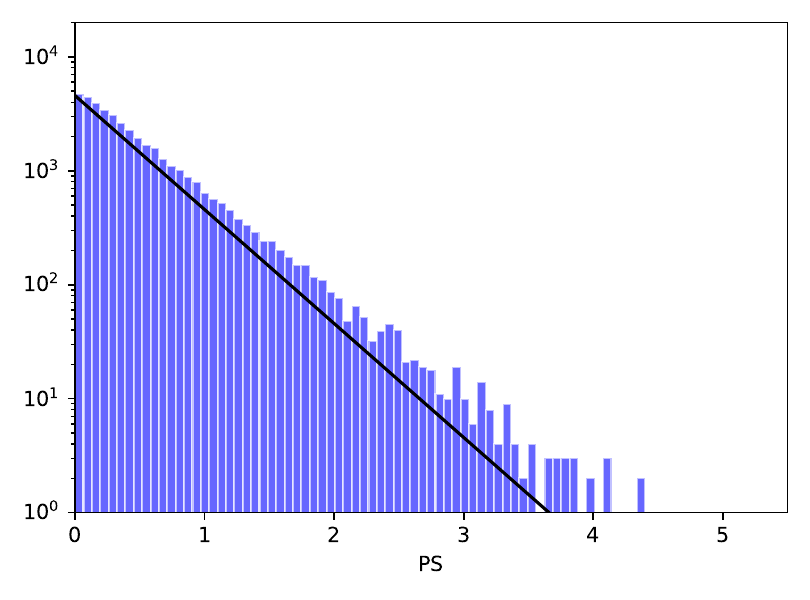}
        \caption{(\textit{Left}): P-value statistic (PS) residual map of the ROI, following the data-model comparison method developed by \cite{Bruel21}. Each map is saturated at $\rm{PS}=4.2$, corresponding to $4\sigma$ for point-like sources (\textit{Right}): PS distribution of each map. The top, middle and bottom panels correspond to the analyses above 10~GeV, 1~GeV and 100~MeV, respectively. If all sources are accounted for in the model, the PS histogram should follow a distribution $\propto 10^{-x}$ (black line). }
        \label{fig:PS_maps}
\end{figure*}

\newpage

\section{Additional excess with respect to 4FGL-DR4}\label{app:sources}

\begin{table*}[h]
\caption{Point-like \textit{Fermi}-LAT excesses found in the analysis above 1 GeV. Those excesses could be either new sources or diffuse emission not accounted for in our Galactic diffuse model.}             
\label{table:etacar}      
\centering                          
\begin{tabular}{c c c c c c}        
\hline\hline      
Source & $(l,b)$ & Offset ($^{\circ}$) & TS & Integrated flux (ph/cm$^2$/s) & $\Gamma$\\  
\hline     
PS J1051.2-5931 & (288.224, -0.127) & 3.4 & 41.5 &  $(8.16 \pm 1.50) \cdot 10^{-10}$ & $2.68\pm0.18$ \\
PS J1111.5-5751$^{\rm{a}}$ & (289.939, +2.463) & 3.4 & 29.2 & $(5.89 \pm 2.41) \cdot 10^{-11}$ & $1.31\pm0.17$   \\
PS J1047.9-5943 & (287.943, -0.495) & 3.7 & 39.7 &  $(8.64 \pm 1.74) \cdot 10^{-10}$ & $2.40\pm0.21^{\rm{b}}$ \\  
PS J1046.1-5950 & (287.793, -0.710) & 3.8 & 90.0 &  $(1.75 \pm 0.28) \cdot 10^{-9}$ & $2.45\pm0.12$ \\
PS J1044.1-5922 & (287.353, -0.417) & 4.3 & 46.6 &  $(1.06 \pm 0.20) \cdot 10^{-9}$ & $2.51\pm0.16$ \\
PS J1043.1-5936 & (287.346, -0.682) & 4.3 & 113.4 &  $(1.97 \pm 0.25) \cdot 10^{-9}$ & $2.61\pm0.11$ \\
PS J1043.7-5851 & (287.060, +0.028) & 4.6 & 26.3 &  $(4.73 \pm 1.16) \cdot 10^{-10}$ & $2.27\pm0.27^{\rm{b}}$ \\ 

\hline  
\end{tabular}
\tablefoot{$^{\rm{a}}$Although we do not discuss associations for the excesses, given the hardness of PS J1111.5-5751 we note that it is spatially consistent with the X-ray sources 1eRASS~J111126.6-575208 and 1eRASS~J111113.9-574927 \citep{Merloni2024}.  $^{\rm{b}}$These excesses were initially detected between $4\sigma$ and $5\sigma$, therefore $\Gamma$ is a fixed parameter in the ROI refit. The error provided is derived during the initial excess localization and not the final fit.}
\label{tab:lat}
\end{table*}

\end{appendix}

\end{document}